\documentclass[prd,onecolumn,
  tightenlines,nofootinbib,eqsecnum,
]{revtex4}

\usepackage[]{babel}

\usepackage{amsmath}
\usepackage{amsfonts}
\usepackage{amssymb}
\usepackage{bm}
\usepackage{hyperref}
\usepackage{mathrsfs}
\usepackage{graphicx}

\usepackage{empheq}

\usepackage{ulem}
\usepackage[usenames]{color}

\definecolor{darkgreen}{rgb}{0,0.5,0}

\hypersetup{
    bookmarks=true,         
    unicode=false,          
    pdftoolbar=true,        
    pdfmenubar=true,        
    pdffitwindow=false,     
    pdfstartview={FitH},    
    pdftitle={My title},    
    pdfauthor={Author},     
    pdfsubject={Subject},   
    pdfcreator={Creator},   
    pdfproducer={Producer}, 
    pdfkeywords={keyword1} {key2} {key3}, 
    pdfnewwindow=true,      
    colorlinks=true,       
    linkcolor=red,          
    citecolor=cyan,        
    filecolor=magenta,      
    urlcolor=darkgreen,           
    linktocpage=true
}

\allowdisplaybreaks

\DeclareSymbolFontAlphabet{\mathrsfs}{rsfs}
\DeclareMathAlphabet{\mathcal}{OMS}{cmsy}{m}{n}

\newcommand{\beq}{\begin{equation}}
\newcommand{\eeq}{\end{equation}} 

\newcommand{\dd}{\mathrm{d}}

\begin{document}

\title{150 ans de Relativité et de Gravitation\\{\small{\rm [Paru dans ``\textit{Les 150 ans de la Soci\'et\'e Fran\c{c}aise de
        Physique}'', \'Edition Diffusion Presse Sciences (2023)]}}}

\author{Luc \textsc{Blanchet}}\email{luc.blanchet@iap.fr}
\affiliation{Institut d'Astrophysique de Paris, UMR 7095 du CNRS,
  Sorbonne Universit{\'e}s \& UPMC Universit\'e Paris
  6,\\ 98\textsuperscript{bis} boulevard Arago, 75014 Paris, France}

\date{\today}

\begin{abstract}
\end{abstract}


\maketitle

\vspace{0.3cm}
\section{La génèse difficile de la relativité restreinte} 
\vspace{0.3cm}

L'année 1873 de la création de la Société Française de Physique est aussi celle de la publication par Maxwell de son Traité d'Electricité et de Magnétisme avec ses fameuses équations de l'électromagnétisme, qui expriment sous forme locale les lois empiriques de Gauss, Faraday et Ampère (avec une modification cruciale\,: l'introduction du ``courant de déplacement''), et fondent la théorie moderne de la lumière, comprise comme une onde de perturbations magnétiques et électriques transverses. La vitesse de la lumière, en notation moderne, est donnée par $c=(\varepsilon_0\mu_0)^{-1/2}$  où $\varepsilon_0$ et $\mu_0$ sont la permittivité diélectrique et la perméabilité magnétique du vide.
\begin{figure}[b]
\begin{center}
\includegraphics[width=5cm]{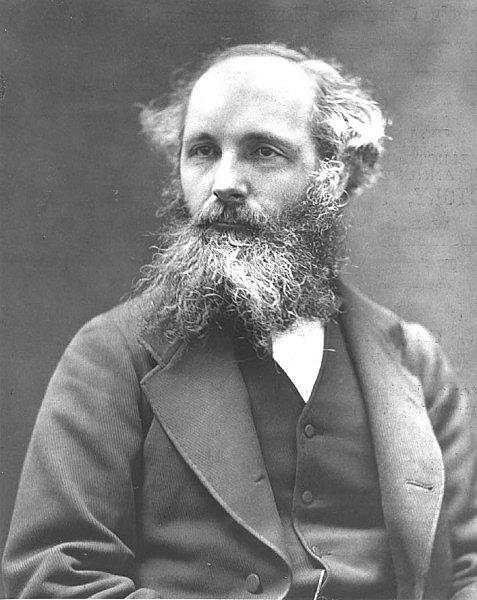}
\end{center}
\caption{James Clerk Maxwell publie en 1873 son Traité d'électromagnétisme.}\label{}
\end{figure}

\`A cette époque la notion de champ se propageant dans le vide était inconcevable, et Maxwell supposait que la lumière est une vibration d'un milieu matériel appelé l'``éther'', reminiscent de l'éther luminifère de Huygens (1678). De plus, les équations n'étant pas invariantes sous le groupe de transformations de Galilée (qui implique en particulier la loi d'addition des vitesses), elles ne devaient être valables que dans le référentiel privilégié de l'éther. Donc des mesures de la vitesse de la lumière en laboratoire devraient permettre de déterminer expérimentalement la vitesse de la Terre par rapport à cet éther\,! Mais la quête du ``vent d'éther'', par la mesure de la vitesse absolue de la Terre, fut infructueuse\,: les expériences d'interférométrie de Michelson et Morley (1887) amenèrent à conclure à l'isotropie de la vitesse de propagation de la lumière, c'est-à-dire la même vitesse dans toutes les directions, et donc l'impossibilité d'observer le mouvement absolu de la Terre et de sentir le souffle du vent d'éther !  

On a du mal à imaginer la confusion qui régnait à cette époque de génèse de la relativité moderne. L'éther était considéré comme un milieu ayant des propriétés mécaniques et élastiques. Dans la théorie de Fresnel (1830) l'éther était partiellement entraîné par les corps en mouvement, avec, mystérieusement, un coefficient d'entraînement égal à $1-1/n^2$ dans le cas de la lumière se propageant dans un milieu transparent d'indice $n$. L'expérience de Fizeau de 1851 mesurant par interférométrie la vitesse de la lumière entraînée par des courants d'eau confirmait la formule d'entraînement. Mais le résultat de cette expérience était clairement en désaccord avec la loi d'addition des vitesses de Galilée dans le cas de la lumière. Et bien sûr le résultat négatif des expériences de Michelson et Morley semblait en contradiction avec l'expérience de Fizeau. 
\begin{figure}[h]
	\begin{center}
		\begin{tabular}{c}
			\hspace{-0.5cm}\includegraphics[width=8cm]{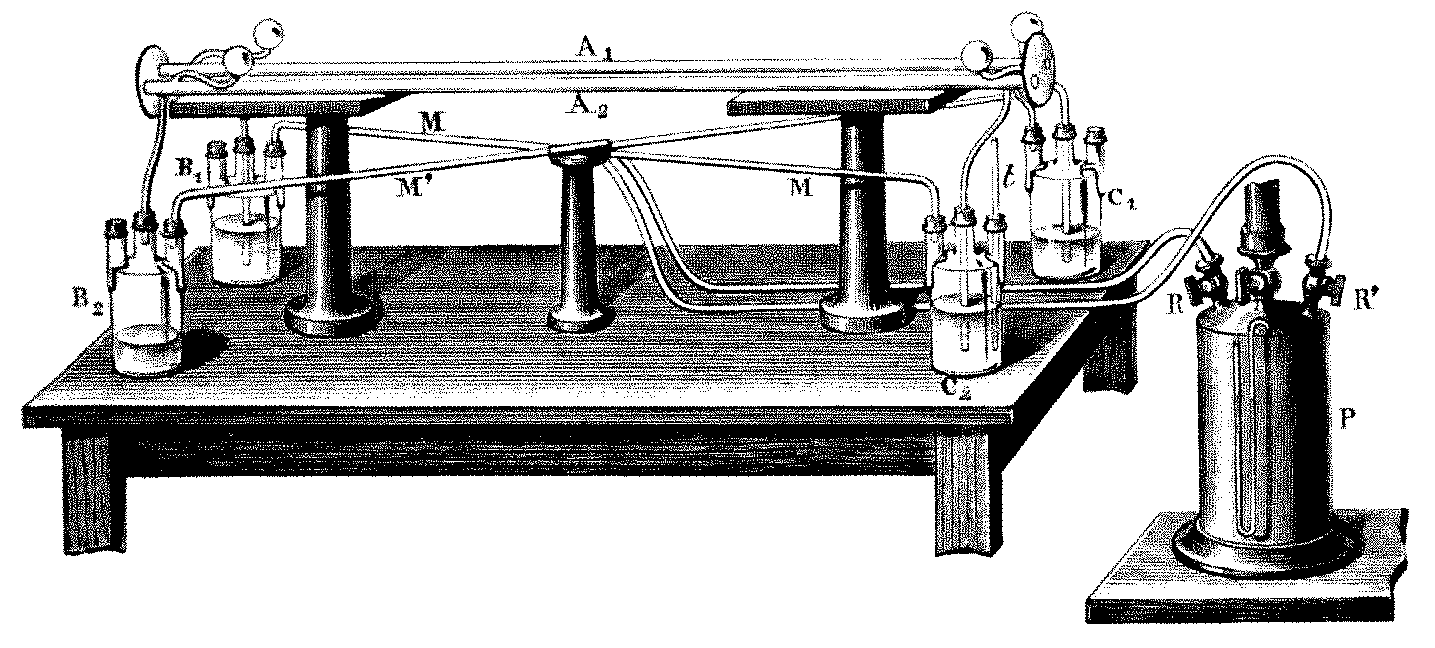}
			\hspace{0.9cm}\includegraphics[width=8cm]{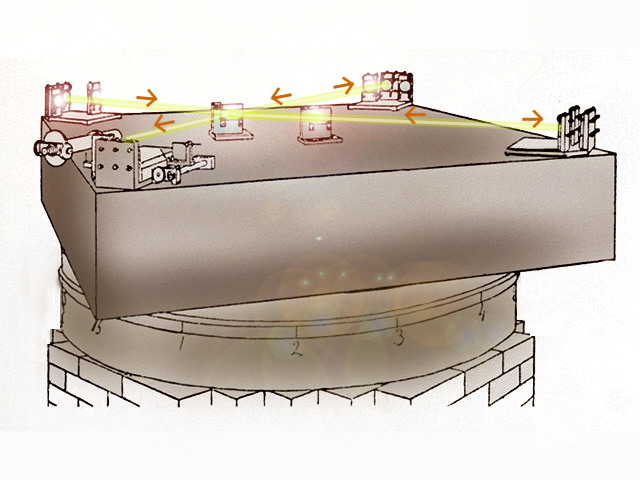}
		\end{tabular}
		\caption{La fameuse expérience de Fizeau (1851) sur la vitesse de la lumière entrainée par un courant d'eau (à gauche), et l'expérience de Michelson et Morley (1887) démontrant l'isotropie de la vitesse de propagation de la lumière (à droite).}
		\label{}
	\end{center}
\end{figure}

Sur le front théorique les progrès survinrent en 1892 et 1895 avec l'idée de Lorentz selon laquelle les charges électriques sont portées par des particules soumises à l'action des champs électriques et magnétiques qui matérialisent l'éther. Lorentz présuppose un éther complètement au repos, mais incorpore l'hypothèse de Fitzgerald (1889) de contraction des longueurs des corps matériels dans la direction du vent d'éther, ce qui lui permet de rendre compte du résultat négatif des expériences d'éther à l'ordre $v/c$. Essayant d'étendre le résultat à tous les ordres en $v/c$ il obtient une forme préliminaire de ce que Poincaré nommera en 1905 ``transformations de Lorentz''. Cependant la loi de transformation pour la densité de charge et de courant était incorrecte et la théorie de Lorentz n'excluait pas complètement la possibilité de détecter l'éther. 
\begin{figure}[b]
	\begin{center}
		\includegraphics[width=8cm]{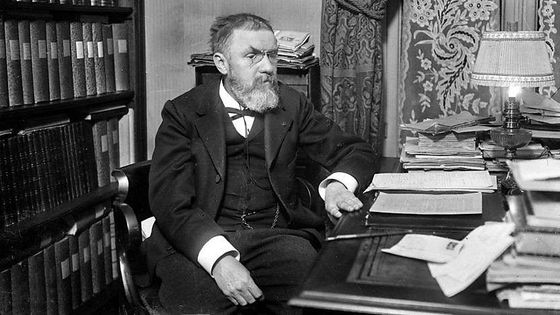}
	\end{center}
	\caption{Henri Poincaré est célèbre (entre autres) pour ses contributions à la relativité et ses remarques prémonitoires sur la ``nouvelle mécanique''. Il fut président de la SFP en 1902.}\label{}
\end{figure}

Poincaré en 1905 corrigea les erreurs dans l'article de Lorentz et incorpora les forces non-électromagnétiques dans ce qu'il appela ``la nouvelle mécanique''. A ce stade les équations de Maxwell étaient devenues parfaitement invariantes sous l'action des transformations de Lorentz\,! Dans son ``mémoire de Palerme'' de 1905, Poincaré montre que les transformations de Lorentz laissent invariante la combinaison $x^2 – c^2t^2$ qui deviendra en 1910 dans les mains de Minkowski l'intervalle ou distance dans l'espace-temps. Résultat crucial pour la physique moderne, il prouve que les transformations de Lorentz forment avec les translations d'espace et de temps un groupe appelé de nos jours groupe de Poincaré [1]. Dans ses cours de 1899 Poincaré avait suggéré que l'on a à faire à un nouveau principe de relativité, et depuis 1889 il jugeait l'éther probablement non nécessaire.

La palme d'or pour la compréhension physique de la nouvelle mécanique revient à Einstein dans son article de 1905 sur l'électrodynamique des corps en mouvements.  En fait Einstein avait tellement ``tout compris'' qu'il ne prendra pas la peine de citer les travaux antérieurs, et il mentionne seulement le nom de Lorentz au détour d'un paragraphe. Dorénavant les transformations de Lorentz-Poincaré remplacent les transformations de Galilée et définissent la nouvelle relativité appelée relativité restreinte. L'invariance sous ces transformations implique les lois de conservation de l'énergie, du moment cinétique, de l'impulsion et de la position du centre de masse. La vitesse de la lumière $c$ est une nouvelle constante fondamentale de la physique. La loi d'addition des vitesses est plus compliquée que celle de Galilée; elle donne $c'=c$ lorsqu'elle est appliquée à la lumière, ce qui rend compte de l'expérience de Michelson-Morley. Elle explique aussi très simplement le résultat de l'expérience de Fizeau comme l'a montré von Laue [2]. La nouvelle dynamique relativiste implique la célèbre relation entre la masse au repos et l'énergie $E=mc^2$. La variable ``commode'' de temps local utilisée par Poincaré n'est autre que le temps physique qui s'écoule dans un référentiel en mouvement, tel que mesuré par la montre-bracelet d'un observateur. 

Et l'éther n'existe pas.\footnote{Le référentiel privilégié du fond diffus cosmologique découvert en 1965 pourrait être vu comme une version moderne de l'éther. La Terre a une vitesse de environ 370 km/s par rapport à ce référentiel cosmologique.}

\section{Découverte de la relativité générale et premiers succès}

Si les lois de l'électromagnétisme étaient maintenant parfaitement invariantes sous la nouvelle relativité, \textit{quid} de la gravitation\,? En 1906 Poincaré proposa la première théorie relativiste de la gravitation, qui consistait à remplacer l'équation de Poisson pour le potentiel newtonien par une équation d'onde de type d'Alembertien, et promouvoir le potentiel newtonien en un champ ``scalaire''. Cette théorie conduit à une notion primitive d'onde gravitationnelle que Poincaré appela ``onde gravifique'' [3].
\begin{figure}[h]
	\begin{center}
		\includegraphics[width=4.5cm]{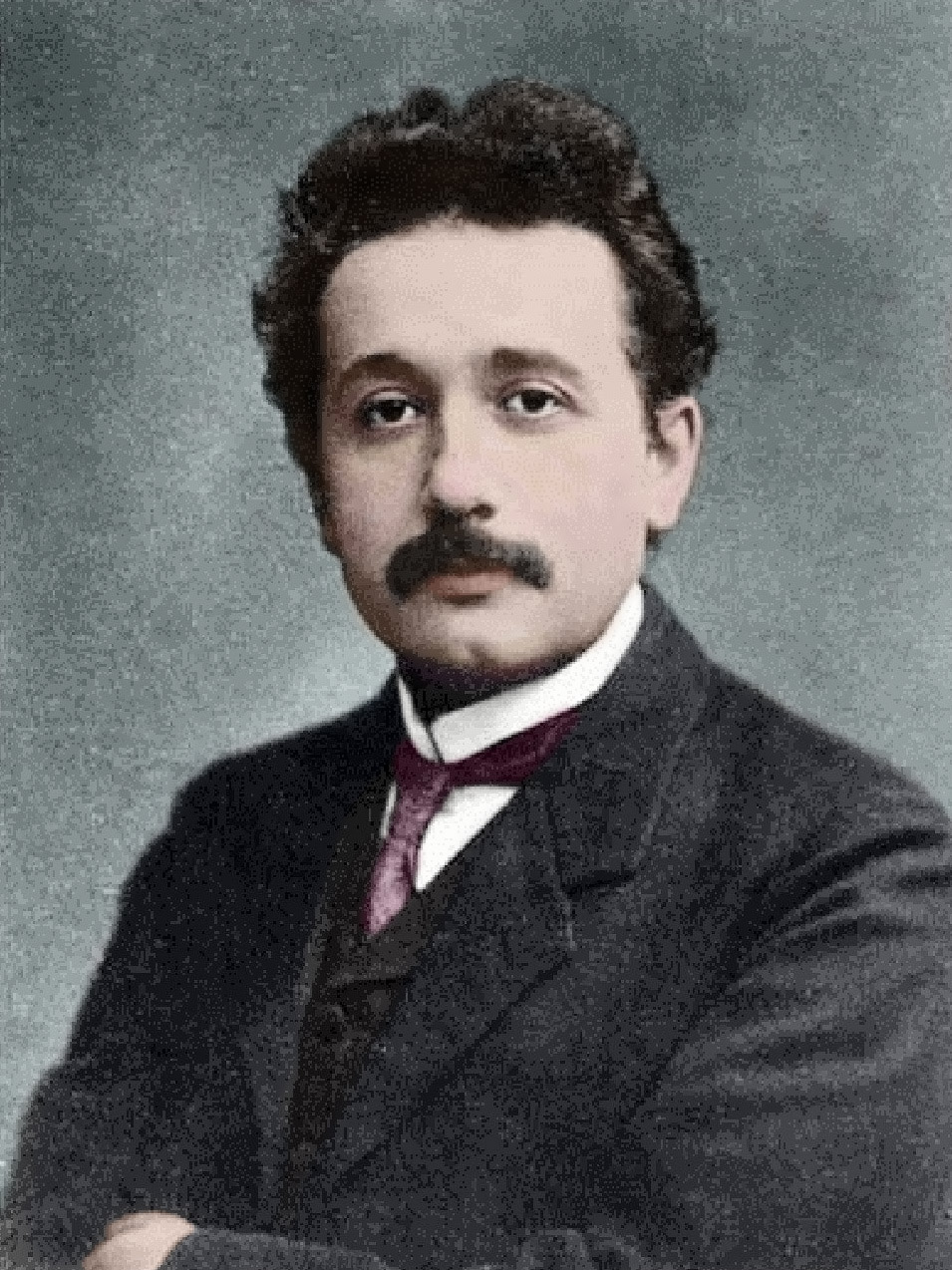}
	\end{center}
	\caption{Einstein crée en 1915 la théorie de la gravitation ou relativité générale, considérée encore aujourd'hui comme la plus belle théorie physique.}\label{}
\end{figure}

Dix années d'intense créativité intellectuelle, unique dans l'histoire des sciences, mèneront Einstein à la formulation de la relativité générale en novembre 1915. Cette théorie, d'une cohérence mathématique et d'une simplicité stupéfiantes, combine le postulat de relativité restreinte avec un fait expérimental propre à la gravitation et qui est érigé en principe fondamental\,: le principe d'équivalence, ou principe d'universalité de la chute des corps, c'est-à-dire indépendamment de leur composition interne, dans un champ de gravitation. Le nom de principe d'équivalence vient de l'équivalence entre les vrais champs gravitationnels, engendrés par des masses, et les champs d'accélération inertielle. 

Sur la base du principe d'équivalence, par une belle expérience de pensée, Einstein prédit en 1911 le décalage de la fréquence d'un photon vers le rouge dans un champ de gravitation par rapport à sa fréquence loin de toute distribution de masse. A cette époque Einstein se convainc que la seule façon d'incorporer le principe d'équivalence est de faire en sorte que la force gravitationnelle soit la manifestation des propriétés géométriques de l'espace-temps. C'est Grossmann, qui fut un ancien camarade d'Einstein au polytechnicum de Zurich, qui mit Einstein sur la voie de la géométrie riemannienne [4], puissante théorie créée par les mathématiciens soixante ans avant son application à la physique\,! Einstein, qui n'avait eu jusque là que peu d'inclinaison pour les mathématiques, se mit à admirer leur efficacité\,: 
\vspace{-0.1cm}
\begin{itemize}
\item[] ``\textit{Comment se fait-il que les mathématiques, étant après tout un produit de la pensée humaine \\indépendante de l'expérience, soient si admirablement adaptées aux objets de la réalité\,?}''
\end{itemize}
\vspace{-0.1cm}
On peut facilement imaginer que cette question poursuivit Einstein sa vie durant.

Des discussions entre Einstein et le mathématicien Hilbert pendant l'été 1915 ressortiront un principe d'action pour le champ gravitationnel, qui est décrit par la ``métrique'' de l'espace-temps. La métrique est un champ tensoriel qui généralise le potentiel newtonien. Comme l'a montré Weyl à partir de 1918 il s'agit d'un champ de jauge de masse nulle et de spin égal à deux. L'action d'Einstein-Hilbert de la relativité générale est d'une simplicité biblique : juste l'intégrale sur tout l'espace-temps du scalaire de courbure, dénoté ``$R$'' en l'honneur de Riemann. Les champs de matière se couplent à la métrique de l'espace-temps ce qui permet d'assurer le principe d'équivalence. Les particules suivent les géodésiques de l'espace-temps. Les théories qui ont cette propriété de coupler la matière à la métrique de l'espace-temps s'appellent théories métriques; la relativité générale est la plus simple des théories métriques.

Dès l'obtention des équations du champ gravitationnel, Einstein s'attelle au problème du mouvement. Il prouve que les corrections relativistes au mouvement d'une planète sur une orbite keplerienne impliquent une précession du grand axe de l'ellipse par rapport au mouvement newtonien. Dans le cas de Mercure la précession relativiste est de $43''$ d'arc par siècle et doit se rajouter à l'effet des perturbations newtoniennes induites par les autres planètes. L'accord avec les observations est parfait et explique un mystère connu depuis les calculs de Le Verrier en 1845. Einstein a raconté qu'il n'en dormit plus d'enthousiasme pendant une semaine\,!

L'effet de précession relativiste est mesuré de nos jours avec une précision de l'ordre de $8 \,10^{-5}$ dans le Système Solaire et dans le mouvement des pulsar binaires. Récemment l'effet a été mesuré pour l'étoile $S_2$ en orbite autour du trou noir supermassif au centre de notre galaxie [5]. Il devrait aussi pouvoir être mesuré dans le cas de l'exoplanète HD80606b [6], qui est située à 217 années-lumière et qui est un cas unique et remarquable avec une orbite très excentrique ($e \simeq 0.93$), ce qui augmente l'effet de précession relativiste, et sans planètes perturbatrices dans ce système, de sorte que l'effet relativiste est dominant.

En 1919 Eddington et Dyson montent deux expéditions, l'une dans l'île de Principe dans le golfe de Guinée et l'autre à Sobral au Brésil, pour photographier lors d'une éclipse solaire les étoiles à proximité du Soleil et ainsi mesurer l'angle de déviation des rayons lumineux par le Soleil. Malgré une grande incertitude sur la mesure, les plaques photographiques sont jugées probantes, et l'angle de déviation de la lumière par le Soleil est mesuré à $1,75$ secondes d'arc. L'effet est deux fois plus important que ce que prévoit la théorie de Newton. La relativité générale est confirmée. Einstein est célèbre\,!
\begin{figure}[h]
	\begin{center}
		\includegraphics[width=4.5cm]{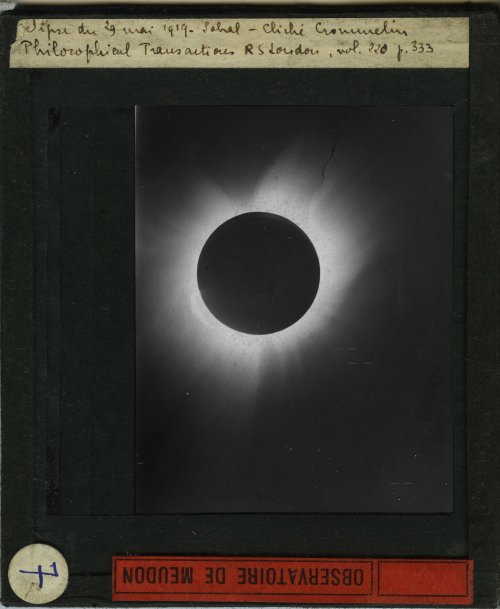}
	\end{center}
	\caption{L'éclipse historique du 29 mai 1919 qui a permis de mettre en évidence l'effet de déviation de la lumière par le Soleil, en accord avec la prédiction d'Einstein.}\label{}
\end{figure}

La façon moderne de quantifier les effets relativistes dans le Système Solaire est d'introduire des paramètres dits ``post-newtonien''\,: $\gamma$ qui quantifie la déviation de la lumière ainsi que l'effet de retard gravitationnel [7], et $\beta$ qui est une mesure de la non-linéarité de la théorie [8]. La mesure de ces paramètres est conforme avec la prédiction de la relativité générale à mieux que $10^{-5}$ pour $\gamma$ et à mieux que $10^{-4}$ pour $\beta$. Les pionniers du schéma d'approximation ``post-newtonien'' en relativité générale, ou approximation de mouvement lent $v/c\to 0$, sont Droste (1916) et Lorentz et Droste (1917) [9], qui ont résolu le problème du mouvement de deux corps à l'approximation $(v/c)^2$. Ces travaux précurseurs sont souvent ignorés par la communauté, qui préfère citer l'article de Einstein, Infeld et Hoffmann de 1939 [10], qui pourtant est moins complet et ``moderne'' que les travaux de 1917. 

La relativité générale, après ses succès initiaux, a été longtemps considérée comme une très belle construction mais avec peu de pertinence au monde réel de la Physique, dominé par la mécanique quantique et plus tard la théorie quantique des champs [11]. Les développements nouveaux qui ont commencé dans les années 1960 (à la fois théoriques, observationnels et expérimentaux), notamment dans le domaine des trous noirs et des ondes gravitationnelles, ont donné à la relativité générale une place centrale dans la physique d'aujourd'hui.

Einstein n'a malheureusement pas connu le renouveau de sa théorie. Il ne croyait pas en l'existence des trous noirs et ne pensait pas que l'on puisse un jour détecter les ondes gravitationnelles. Il ne semble pas avoir anticipé les propriétés des ondes gravitationnelles qui sont en train de conduire à une révolution en Astronomie\,! Mais nul doute que Einstein aurait adhéré à la nouvelle physique et serait devenu un fervent ``relativiste''.

\section{Le principe d'équivalence aujourd'hui}

La relativité générale a été bâtie de façon à satisfaire exactement (presque par définition) le principe d'équivalence. Notons que ce principe est plutôt non-naturel dans le sens où il rend la théorie de la gravitation très différente de celle décrivant les autres interactions. Ce principe ne constitue pas une symétrie fondamentale de la Physique, au même titre par exemple que le principe d'invariance locale de jauge en physique des particules. Einstein lui-même l'a initialement appelé l'``hypothèse d'équivalence'' avant de l'ériger en un ``Grand Principe''. Le principe d'équivalence d'Einstein comprend non seulement l'universalité de la chute libre des corps dans le champ gravitationnel, mais aussi le fait que dans le référentiel qui accompagne le mouvement de chute libre des corps (et donc dans lequel les corps sont en état d'``apesanteur''), on retrouve la relativité restreinte et tout ce que l'on connaît de la physique non-gravitationnelle. 

L'expérience historique sur le principe d'équivalence est celle d'Eötvös [12] qui utilise un pendule de torsion sur lequel sont suspendus deux corps de compositions différentes. Les deux corps sont soumis au champ gravitationnel de la Terre et au champ d'accélération inertielle de rotation de la Terre sur elle-même. Une non-équivalence entre le champ gravitationnel et le champ inertiel se traduit par une réponse différente des corps selon leur rapport masse inertielle à masse gravitationnelle et l'apparition d'un couple de torsion mesurable. Les versions modernes de cette expérience sont basées sur le champ inertiel de rotation de la Terre autour du Soleil [13].
\begin{figure}[h]
	\begin{center}
		\begin{tabular}{c}
			\hspace{-0.8cm}\includegraphics[width=5.5cm]{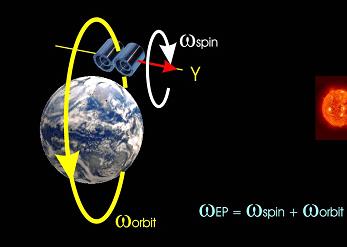}
			\hspace{1.8cm}\includegraphics[width=5.0cm]{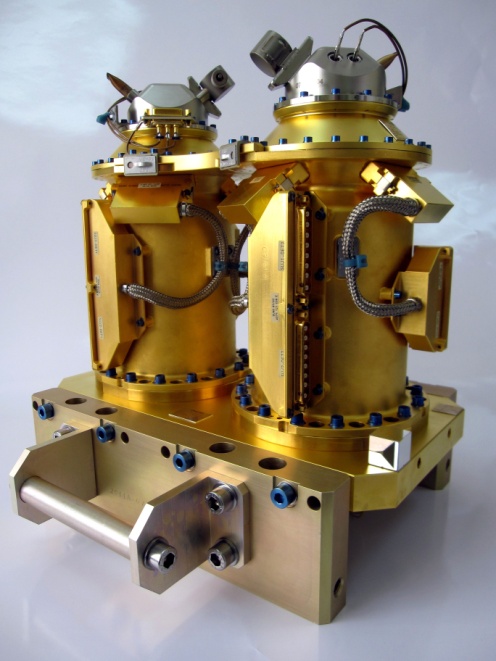}
		\end{tabular}
		\caption{Le satellite Microscope a testé le principe d'équivalence en orbite terrestre au niveau record $10^{-15}$; à droite les accéléromètres électrostatiques de l'ONERA à bord de Microscope.}
		\label{}
	\end{center}
\end{figure}

Récemment l'expérience française Microscope (Touboul et al. [14]) a permis de battre le record de précision sur le test du principe d'équivalence en embarquant un accéléromètre construit par l'ONERA sur un satellite en orbite terrestre conçu par le CNES. On mesure l'accélération relative de deux cylindres emboités de composition différente (l'un en platine l'autre en titane) dans le champ de la Terre.\footnote{Un deuxième accéléromètre fonctionne avec des corps de composition identique et permet de s'assurer que la violation éventuelle du principe d'équivalence n'agit pas dans ce cas.} La précision obtenue est égale à $10^{-15}$ sur le paramètre d'Eötvös qui est l'écart d'accélération relative entre les deux matériaux.    

Dans une autre classe de tests du principe d'équivalence on utilise l'interférométrie atomique dans le champ de gravitation de la Terre, ce qui conduit à une mesure de l'accélération de l'atome que l'on compare avec la chute libre d'un corps macroscopique ``classique''. On peut utiliser aussi deux espèces différentes d'atomes (ou deux isotopes différents d'un même atome) dans un même interféromètre atomique. Ces tests peuvent être qualifiés de ``quantique'' par opposition au test de Microscope car c'est bien sûr la fonction d'onde quantique de l'atome qui produit les interférences.    

Le meilleur test de l'isotropie de la vitesse de la lumière (qui est l'une des facettes de l'invariance locale de Lorentz et donc du principe d'équivalence d'Einstein) a été obtenu dans une expérience de type Michelson-Morley par Brillet et Hall avec la précision de $10^{-15}$ [15]. D'autres tests de l'invariance locale de Lorentz incluent des comparaisons de la fréquence d'horloges de type différents (fonctionnant avec des atomes différents) lorsque la vitesse et/ou l'orientation du laboratoire est changée, par exemple à cause de la rotation et du mouvement de la Terre sur son orbite. 
\begin{figure}[h]
	\begin{center}
		\includegraphics[width=5cm]{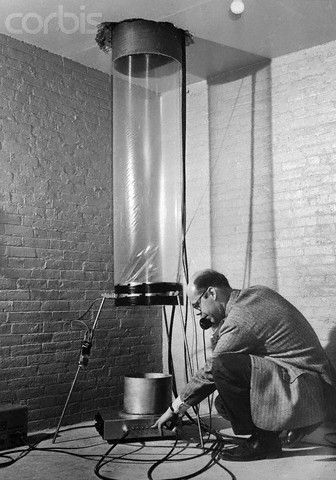}
	\end{center}
	\caption{L'expérience de Pound et Rebka. Un échantillon solide contenant du fer 57 émettant des photons $\gamma$ est placé au sommet d'une tour de 22,5 mètres de hauteur. Les photons sont absorbées par un autre échantillon de fer 57 placé à la base de la tour. Un décalage Doppler est introduit artificiellement par le mouvement de l'échantillon récepteur, ce qui compense le décalage gravitationnel et constitue le signal observable.}\label{}
\end{figure}

Quant à l'effet Einstein ou décalage gravitationnel des fréquences, qui est aussi l'une des facettes du principe d'équivalence, il fut mesuré par Pound et Rebka en 1960 [16], ce qui a constitué la première vérification précise en laboratoire de la relativité générale. L'effet Einstein se manifeste de façon équivalente dans une expérience d'horloges, où on compare les battements d'horloges à différents points du potentiel gravitationnel, et on vérifie que les différences de fréquences d'horloges ne dépendent que de la différence des potentiels gravitationnels, et non de la nature et de la structure interne des horloges. En 1976 un maser à hydrogène était embarqué sur une fusée à une altitude de 10.000 km et sa fréquence comparée grâce à un lien $\mu$-onde à une horloge  placée au sol, ce qui a permis un test du décalage vers le rouge avec une précision de $10^{-4}$ [17]. L'agence spatiale européenne va placer en 2026 sur la station spatiale internationale l'expérience ACES (Atomic Clock Ensemble in Space), qui comprendra l'horloge ultra-stable à atomes refroidis par laser PHARAO, et va tester, en plus de nombreuses autres applications en physique fondamentale et en métrologie, le décalage gravitationnel avec une précision d'environ $10^{-6}$ [18]. 

On ne peut quitter ce sujet sans parler de la version forte du principe d'équivalence. Celle-ci stipule que l'universalité de la chute libre des corps s'applique même pour des corps avec une auto-gravité importante comme dans le cas des planètes. Dans le repère en chute libre qui accompagne les corps accélérés par un champ gravitationnel ``extérieur'', non seulement on retrouve la relativité restreinte mais on peut faire une expérience gravitationnelle comme l'expérience de Cavendish (1798) sur la mesure de la constante de Newton, et trouver le même résultat qu'en l'absence du champ gravitationnel ``extérieur''\,!

On voit bien que ce principe implique une contrainte très forte sur la théorie. Pas étonnant qu'il n'existe que deux théories qui satisfont au principe d'équivalence fort\,: la relativité générale et la théorie scalaire de Nordström (1911) qui est un ancêtre de la relativité générale par ailleurs invalidé par le test de la déviation de la lumière par le Soleil et l'émission d'ondes gravitationnelles par les pulsars binaires.

La Terre ayant une gravitation interne beaucoup plus importante que la Lune, et toutes deux étant en ``chute libre'' dans le champ de gravitation du Soleil, on peut tester le principe d'équivalence fort en recherchant une anomalie dans la distance Terre-Lune. Celle-ci est mesurée avec une précision de l'ordre du millimètre par des tirs laser sur la Lune (notamment depuis le plateau de Calern, près de Grasse) et on en déduit un test du principe d'équivalence fort au niveau $10^{-12}$ [19].

Aujourd'hui les tests sur le principe d'équivalence sont considérés comme une voie d'accès à de la physique nouvelle à l'interface entre la relativité générale et les autres interactions fondamentales. On s'attend à ce que la gravitation s'unifie aux autres interactions à l'énergie de Planck $\sim 10^{19} \,\text{Gev}$. Mais la relativité générale souffre d'un grave problème dû à la dimensionnalité de sa constante de couplage (la constante de Newton $G$), qui fait qu'elle n'est pas quantifiable perturbativement et décrite par une théorie des champs renormalisable. Les tentatives d'unification de la relativité générale avec les autres interactions conduisent à des champs nouveaux se superposant au champ tensoriel de la relativité générale, avec comme conséquence observable une violation du principe d'équivalence d'Einstein.

\section{Les trous noirs\,: doutes puis preuves en Astronomie\,!}

Un raisonnement heuristique de Michell (1784) et Laplace (1798) conduit naturellement à la notion de rayon gravitationnel. Si la lumière est faite de corpuscules de vitesse $c$ et de masse $m>0$, alors celle-ci ne peut s'échapper d'un objet de masse $M$ et de rayon $r$ que si l'énergie cinétique des corpuscules est supérieure à leur énergie potentielle gravitationnelle, soit, en utilisant un raisonnement newtonien, $\frac{1}{2} m c^2 > G M m/r$ et donc si le rayon de l'objet est supérieur à un rayon critique $r_g$ donné par $r_g = 2GM/c^2$. La masse du photon disparaît dans ce calcul, c'est le principe d'équivalence\,! A l'inverse, tous les objets ayant $r < r_g$ maintiennent la lumière ``piégée'' dans leur champ de gravitation, et sont donc invisibles pour un observateur à l'infini.
\begin{figure}[h]
	\begin{center}
		\includegraphics[width=5cm]{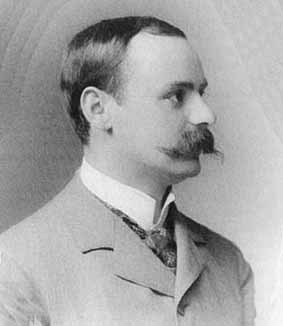}
	\end{center}
	\caption{Karl Schwarzschild découvre en décembre 1915 la solution qui porte son nom alors qu'il est artilleur sur le front russe (il décèdera de maladie en mai 1916).}\label{}
\end{figure}

Le trou noir de Schwarzschild (1916) a été la première solution ``exacte'' des équations d'Einstein\,: c'est une solution valable dans le vide extérieur d'une étoile à symétrie sphérique. Lorsqu'on l'étend formellement dans une région intérieure vide, on trouve une surface critique, l'horizon qui a un rayon égal au rayon gravitationnel de Michell et Laplace. Un théorème fameux de relativité générale dû à Birkhoff (1923), affirme que le champ extérieur d'une étoile en effondrement gravitationnel, tout en maintenant sa symétrie sphérique, est donné par la solution de Schwarzschild. Mais, est-il possible qu'il existe des étoiles dont le rayon soit inférieur au rayon gravitationnel\,?

\`A la densité ordinaire $\rho \sim 1\,\text{g/cm}^3$, de tels objets auraient la masse énorme $M \gtrsim 10^8 \,M_\odot$ calculée par Laplace, et de tels objets de masse $M \sim 1 M_\odot$ auraient la densité énorme $\rho \gtrsim 10^{16} \,\text{g/cm}^3$.

Tous les objets connus dans l'Univers ayant des densités ``ordinaires'' ($\rho\sim 1\,\text{g/cm}^3$) sont maintenus en équilibre contre leur ``poids'' gravitationnel soit par des forces de répulsion d'origine électromagnétique — c'est le cas des objets terrestres, des planètes, etc. — soit par pression de la radiation issue de réactions nucléaires au centre de l'objet — ce sont les étoiles. On peut montrer que ces deux catégories d'objets, ayant des densités ordinaires, ne peuvent jamais dépasser une certaine masse limite qui est bien en-dessous de la masse $\sim 10^8 M_\odot$  de Laplace.

Dans le cas des objets maintenus en équilibre par répulsion électromagnétique, la masse limite, dite masse de Fowler, est de l'ordre de grandeur de la masse de Jupiter ($\sim 10^{-3} M_\odot$). Au delà de la masse de Fowler, les objets maintenus électromagnétiquement deviennent des étoiles, c'est-à-dire que des réactions nucléaires s'allument en leur centre. Dans le cas des étoiles, la masse limite, dite masse de Eddington est de l'ordre de $10^2$ à $10^3$ $M_\odot$. Au delà de la masse de Eddington, les étoiles deviennent instables et leurs couches externes sont ``soufflées'' par la pression de radiation trop intense venant du centre. On pense donc qu'il ne peut exister dans l'Univers d'objets de densité ordinaire ayant une masse $\sim 10^8 \,M_\odot$ et dont le rayon soit inférieur au rayon gravitationnel. Cela ne pourrait être possible que pour des objets de très grande densité.
\begin{figure}[h]
	\begin{center}
		\includegraphics[width=6cm]{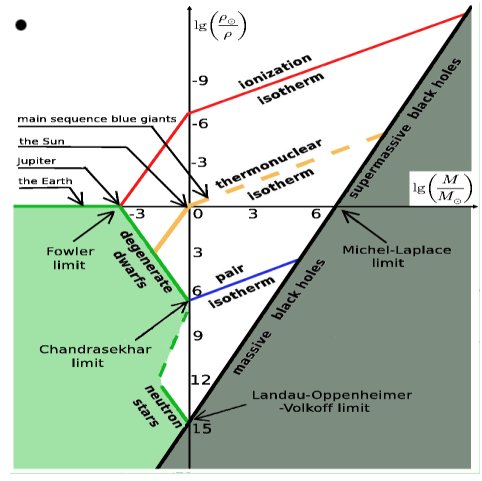}
	\end{center}
	\caption{Le diagramme masse-densité des étoiles (échelle logarithmique). L'effondrement gravitationnel correspond à une trajectoire verticale descendante dans ce diagramme.}\label{}
\end{figure}

Que se passe-t-il lorsque dans une étoile de masse $M \sim 1 M_\odot$ les réactions nucléaires (essentiellement de type hydrogène $\rightarrow$ hélium) s'arrêtent faute d'hydrogène\,? Se peut-il que l'étoile s'effondre jusqu'à atteindre la densité de $10^{16}\,\text{g/cm}^3$ ? En effet, la théorie astrophysique prédit que si l'étoile a une masse supérieure à la masse de Oppenheimer et Volkoff (1939), dont la valeur moderne est de l'ordre de 2 à 3 $M_\odot$, alors l'étoile s'effondre sur elle-même jusqu'à former une étoile à neutrons de rayon $r \sim 10 \,\text{km}$ et de densité $\rho \sim 10^{15}\,\text{g/cm}^3$. Et que, au-delà de cette masse critique, rien ne peut plus arrêter l'effondrement gravitationnel (Oppenheimer et Snyder 1939)\,! L'étoile doit atteindre et dépasser la densité $\rho \sim 10^{16}\,\text{g/cm}^3$ où son rayon devient inférieur à son rayon gravitationnel, et donc finalement être décrite par la solution de Schwarzschild, avec toute la masse de l'étoile concentrée en un point à $r = 0$ -- la singularité où la courbure de l'espace-temps est infinie\,!

Dans les années 1930 les doutes prévalaient sur le scénario précédent\,: Une étoile réelle n'est jamais parfaitement sphérique ! Les petites déviations à la sphéricité n'allaient-elles pas s'amplifier au cours de l'effondrement gravitationnel et changer qualitativement l'état final de l'étoile et en particulier empêcher la formation du trou noir\,? L'état de l'art sur ce problème changea radicalement en 1965 avec le théorème de Penrose. Celui-ci ne repose pas sur l'hypothèse de sphéricité et montre que lorsqu'une étoile s'effondre les déviations à la symétrie sphérique comme la déformation quadripolaire sont rayonnées sous forme d'ondes gravitationnelles. L'état final de l'effondrement est bel et bien un trou noir, conséquence inéluctable de la relativité générale.

Selon le ``théorème de la calvitie'' le trou noir le plus général est décrit par la solution de Kerr-Newman qui dépend uniquement de trois paramètres: sa masse, le moment cinétique ou spin, et la charge électrique [20]. Pour les trous noirs en astrophysique on peut négliger la charge et il ne reste plus que deux ``cheveux'' sur la ``tête'' du trou noir (selon l'aphorisme de Wheeler)\,: la masse et le spin. Le trou noir en rotation non chargé de Kerr [21] est l'état final de l'effondrement d'une étoile.
\begin{figure}[h]
	\begin{center}
		\includegraphics[width=4.5cm]{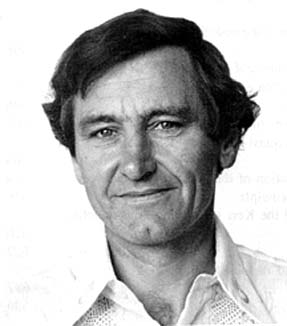}
	\end{center}
	\caption{C'est en 1963 que Roy Kerr découvre la solution pour le trou noir en rotation. L'élucidation des propriétés étonnantes des trous noirs culmineront avec le mécanisme d'extraction d'énergie de Penrose (1969), la notion d'entropie identifiée à la surface de l'horizon par Bekenstein (1973), le rayonnement de Hawking (1974) et les lois de la thermodynamique des trous noirs (Bardeen, Carter et Hawking 1973).}\label{}
\end{figure}

En 1971 des observations rendues possibles par l'essor de la radioastronomie et de l'astronomie X ont commencé à accréditer l'idée qu'un objet invisible en orbite autour d'une étoile supergéante bleue dans notre galaxie, qui a une masse supérieure à la masse limite autorisée pour une étoile à neutron et accrète de la matière chaude provenant de l'étoile en émettant en X, devait être un trou noir. Cette source appelée Cygnus X1 est l'une des meilleures évidences de l'existence d'une population de trous noirs d'origine stellaire dans notre galaxie, c'est-à-dire qui sont issus de l'effondrement gravitationnel suite à la supernova d'une étoile massive.

Dans les années 1990-2000 Genzel et Ghez sont les premiers à apporter des preuves directes de l'existence d'un trou noir supermassif au centre de notre Voie Lactée, à l'emplacement de la radio source Sagittarius $A^\star$. Ils découvrent des étoiles très proches de l'objet central comme la fameuse étoile $S_2$, dont l'orbite elliptique ne peut être expliquée que par la présence d'un trou noir de masse $4\,10^6\,M_\odot$  au foyer de l'ellipse. La rotation du trou noir central n'a pas encore été mesurée de façon certaine, car elle dépend d'un modèle précis pour le disque d'accrétion autour du trou noir qui rende compte des observations en X et en UV. La mesure du spin du trou noir est l'un des buts de l'expérience GRAVITY qui observe avec une précision inégalée la région de Sagittarius $A^\star$, en utilisant le VLT (Very Large Télescope) en mode interférométrique [22]. 

De nos jours on sait que toutes les grandes galaxies recèlent un trou noir géant en leur sein. Des lois d'échelle empiriques relient la masse du trou noir avec la taille du bulbe de la galaxie. Il reste que le mécanisme de formation de ces trous noirs supermassifs est très largement inconnu\,! L'un des scénario envisagés est l'accrétion de matière très rapide par un trou noir initialement d'origine stellaire, dans les premiers temps de la formation des étoiles, lors de la période dite de réionisation à un redshift cosmologique de l'ordre de 10.
\begin{figure}[h]
	\begin{center}
		\includegraphics[width=7cm]{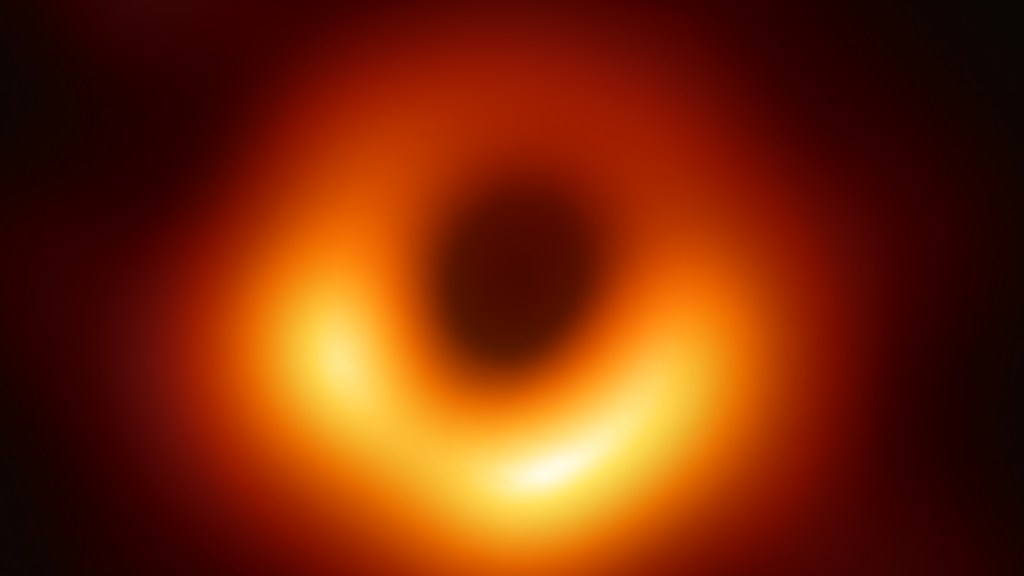}
	\end{center}
	\caption{L'ombre du trou noir supermassif au cœur de la galaxie elliptique Messier 87, imagée par Event Horizon Telescope.}\label{}
\end{figure}

En 2019 la collaboration EHT (Event Horizon Telescope) obtient à l'issue d'un énorme processus de traitement de données la première image d'un trou noir\,: celle du trou noir supermassif au centre de la galaxie elliptique géante M87 située dans l'amas de la Vierge, à une distance de $15,4\,\text{Mpc}$. Ce trou noir a une masse de $6,5\,10^9\,M_\odot$\,! En 2022 EHT a obtenu l'image du trou noir central de la Voie lactée Sagittarius $A^\star$.

Les ondes gravitationnelles vont apporter des preuves encore plus directes de l'existence des trous noirs et de leurs propriétés spécifiques en relativité générale.

\section{Le triomphe des ondes gravitationnelles}

En 1916 Einstein prédit l'existence des ondes gravitationnelles en montrant que la toute nouvelle relativité générale admet des solutions d'ondes [23]. Quelques erreurs techniques émaillent son article, qu'il corrige en 1918 dans un second article [24], où surtout il établit la célèbre formule du ``quadrupôle'', qui donne l'énergie émise sous forme d'ondes gravitationnelles par une source de matière dans l'approximation newtonienne.\footnote{Mais là aussi... Einstein fait une erreur de calcul, la formule du quadrupôle est fausse d'un facteur deux\,!}

Le fait que les ondes gravitationnelles soient quadripolaires est une conséquence du principe d'équivalence qui implique que les moments monopolaire et dipolaires (masse, moment cinétique et position du centre de masse) sont conservés de part les lois du mouvement. Le calcul d'Einstein s'applique à une source mue par des forces non gravitationnelles, mais une preuve que la formule reste valable dans le cas d'une source auto-gravitante, par exemple un système binaire d'étoiles, apparaît dans les premières éditions du traité ``Théorie classique des champs'' de Landau et Lifshitz (1941).

Le pulsar binaire PSR 1913+16 découvert en 1974 par Hulse et Taylor [25] va apporter la preuve observationnelle de l'existence des ondes gravitationnelles et de la validité de la formule du quadrupôle. Le système double formé par le pulsar en orbite autour d'une autre étoile à neutrons pert de l'énergie par émission d'ondes gravitationnelles, ce qui se traduit par une lente dérive de la période orbitale du mouvement et le rapprochement des deux étoiles l'une de l'autre. L'effet est tout petit: $\dd P/\dd t = - 2,4\,10^{-12} \,\text{s}/\text{s}$, mais mesurable [26]\,! L'application de la formule du quadrupôle à un système de deux masses ponctuelles en mouvement sur une ellipse keplerienne [27], et les travaux de Damour et Deruelle sur les équations du mouvement de corps compacts [28], expliquent parfaitement cet effet. C'est un résultat remarquable qui mit en émoi la communauté des relativistes de l'époque.

Dans les années 1960 Weber conçut et construisit le premier détecteur d'ondes gravitationnelles. Il s'agissait d'un cylindre métallique résonnant, dont les oscillations mécaniques induites par le passage d'une onde gravitationnelle sont converties en signal électrique par un transducteur. En 1974 Bonazzola construisit un détecteur similaire à l'observatoire de Meudon. Malheureusement la sensibilité de ces ``barres de Weber'' n'était pas suffisante.

Les détecteurs actuels sont des interféromètres laser de Michelson-Morley, dont les bras de plusieurs kilomètres de long sont constitués de cavités laser résonnantes de Fabry-Perot (1899). La lame séparatrice et les miroirs d'entrée et de sortie des cavités, sont suspendus à un système pendulaire qui permet de s'affranchir des vibrations sismiques terrestres aux basses fréquences. L'onde gravitationnelle est détectée par la variation de la différence de chemin optique entre les deux bras. Outre le bruit sismique, les principales sources de bruit dans l'interféromètre sont le bruit thermique des miroirs et le ``bruit de grenaille'' des lasers. Le grand intérêt de l'interféromètre laser est sa large bande de fréquence au sol, depuis environ 30 Hz jusqu'à quelques kHz.

Le 14 septembre 2015 les détecteurs interférométriques LIGO ont observé le signal de la coalescence de deux trous noirs stellaires de masses 36 et 29 $M_\odot$, à une distance de $400\,\text{Mpc}$ [29]. Cet événement appelé GW150914, a duré une fraction de seconde, durant laquelle les deux trous noirs ont parcouru les derniers cycles orbitaux à une vitesse proche de la lumière avant de fusionner. La fréquence du signal au moment de la fusion est de $150\,\text{Hz}$, proche du maximum de sensibilité des détecteurs. Le signal a été détecté avec un fort signal-sur-bruit de 23 et un taux de fausse alarme équivalent à un niveau de confiance élevé ($5 \sigma$). Le signal est arrivé dans les deux détecteurs LIGO (situés sur les côtes ouest et est américaines), avec un écart de $6,5\,\text{ms}$, compatible avec la distance lumière $10\,\text{ms}$ entre les deux détecteurs, ce qui permet de localiser l'événement dans une portion d'arc de cercle dans le ciel d'environ 600 degrés carrés. A l'époque, le détecteur franco-italien Virgo n'était pas encore en fonctionnement, et la précision sur la localisation de la source n'était pas très bonne. Le 14 août 2017 un événement de trous noirs était observé cette fois avec les trois détecteurs LIGO et Virgo simultanément. La boite d'erreur sur la localisation dans le ciel a donc été drastiquement réduite, environ 30 degrés carrés\,! 
\begin{figure}[h]
	\begin{center}
		\includegraphics[width=12cm]{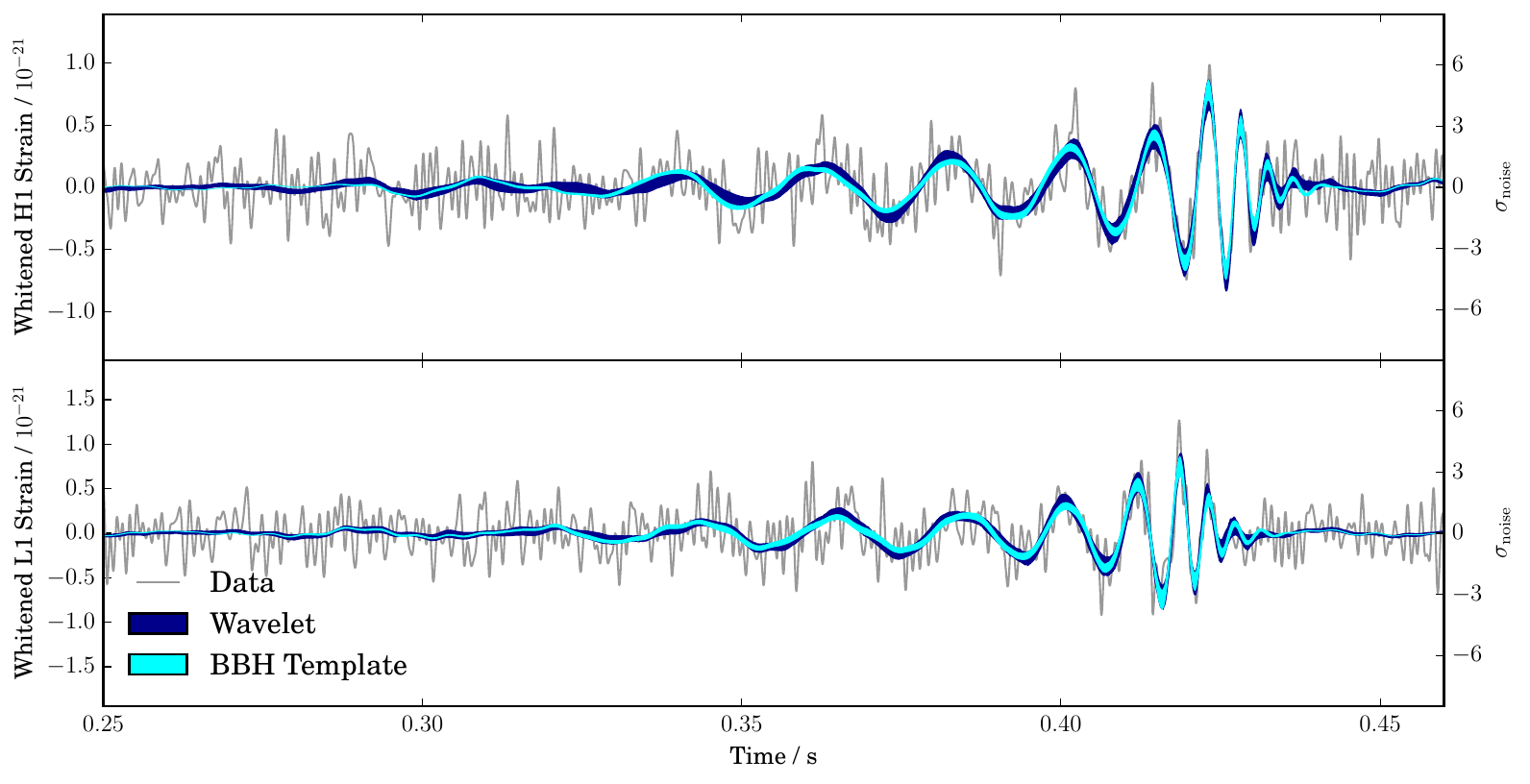}
	\end{center}
	\caption{La première fusion de deux trous noirs détectée\,: GW150914, le 15 septembre 2015. En haut, le signal observé par LIGO/Hanford, en bas par LIGO/Livingstone. On indique les meilleurs ajustements des signaux observés par des calculs numériques en relativité générale, et par une forme d'onde basée sur une superposition d'ondelettes (modèle non physique dans ce cas).}\label{}
\end{figure}

Les masses mesurées des trous noirs sont élevées, en général bien supérieures à celles des trous noirs connus dans notre galaxie comme  Cygnus X1. Expliquer la formation de trous noirs aussi massifs par l'explosion d'une étoile en fin d'évolution stellaire, puis trouver un mécanisme qui conduit à une binaire de deux trous noirs assez serrée pour évoluer par rayonnement gravitationnel dans moins d'un temps de Hubble, représentent des défis pour l'astrophysique non encore résolus actuellement. 

Lors de la fusion des trous noirs, une partie de la masse-énergie du système, prise sur l'énergie de liaison gravitationnelle, est transférée à l'onde. Lors du premier événement de 2015, trois masses solaires ont ainsi été émises sous forme d'onde gravitationnelle. La puissance de l'évènement est d'environ $10^{49}\,\text{W}$, soit l'équivalent de l'énergie totale de 1000 explosions de supernovae émise en une fraction de seconde. On s'attend à ce que la puissance gravitationnelle maximale que puisse émettre une source (avec un rendement de un) soit de l'ordre de la puissance ``de Planck'' $c^5/G \sim 3\,10^{52}\,\text{W}$. On voit donc qu'en termes relativistes la puissance émise est en fait relativement faible. De plus les systèmes de trous noirs n'émettent aucune autre forme d'énergie, car le trou noir est une solution des équations d'Einstein du vide.

Avant la fusion les deux trous noirs suivent une trajectoire en ``spirale rentrante'' à cause de la perte d'énergie due à l'émission des ondes gravitationnelles. La phase spiralante est suivie par la fusion et la formation d'un trou noir unique, initialement fortement déformé à cause de la dynamique de la collision. Le trou noir se relaxe en émettant des ondes gravitationnelles dans les modes dits ``quasi-normaux'', qui sont caractéristiques des modes de résonance du trou noir. La relaxation conduit au trou noir de Kerr, décrit seulement par sa masse et son spin. Les ``patrons d'onde'' utilisés dans l'analyse du signal des détecteurs sont issus de la combinaison des travaux post-newtoniens de Blanchet [30], et Buonanno et Damour [31] avec les calculs de ``relativité numérique''. 

Les observations ont montré que le spin du trou noir final est élevé, environ 0,7 en unité relativiste où le spin maximal est un. Ce résultat est en accord avec la prédiction de la relativité générale. Le spin final est évidemment dû au moment cinétique orbital du système binaire avant la collision, et aux spins individuels des deux trous noirs, mais il se trouve que les spins des trous noirs avant la collision sont faibles.  
\begin{figure}[h]
	\begin{center}
		\includegraphics[width=10cm]{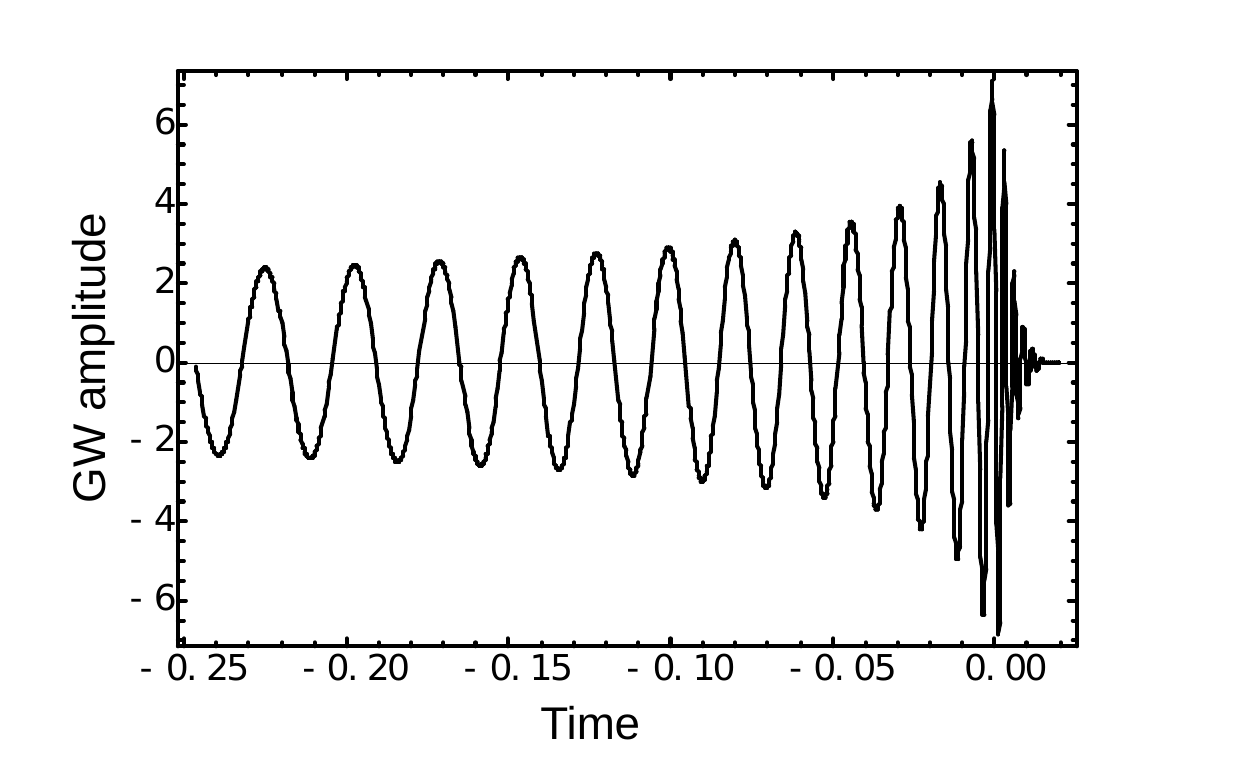}
	\end{center}
	\caption{Le gazouillis gravitationnel des trous noirs binaires. On distingue trois phases successives dans le processus de coalescence des deux trous noirs\,: la phase initiale spiralante, la phase de fusion ou l'amplitude de l'onde atteint son maximal, et la phase de relaxation vers le trou noir final de Kerr.}\label{}
\end{figure}

\section{Astronomie multi-messagère avec les ondes gravitationnelles}  

\begin{figure}[h]
	\begin{center}
		\includegraphics[width=11cm]{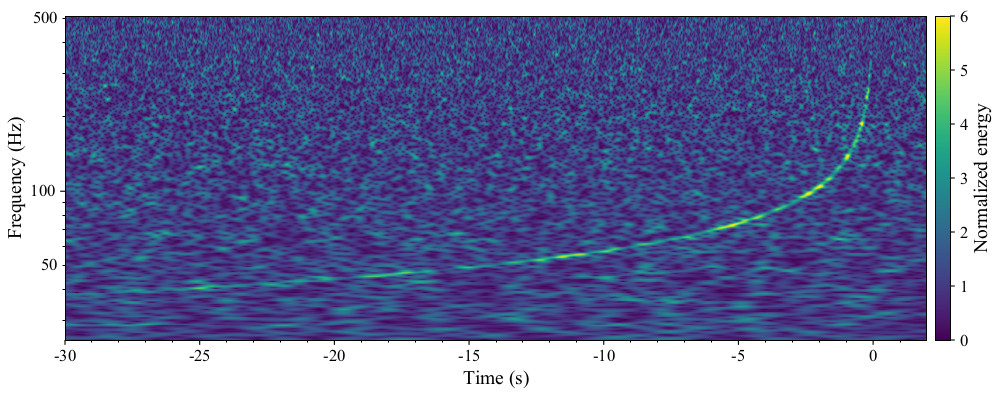}
	\end{center}
	\caption{Le gazouillis gravitationnel des étoiles à neutrons dans un diagramme temps-fréquence.}\label{}
\end{figure}
Le 17 août 2017, les interféromètres LIGO et Virgo détectent un signal d'ondes gravitationnelles de nature différente\,: GW170817, rapidement interprété comme provenant de la fusion de deux étoiles à neutrons [32]. Initialement détecté par une méthode d'analyse temps-fréquence, le signal montre la forme caractéristique du ``chirp'' ou ``gazouillis'' gravitationnel, c'est-à-dire l'augmentation de la fréquence due à la perte d'énergie par ondes gravitationnelles.
\begin{figure}[b]
	\begin{center}
		\begin{tabular}{c}
			\hspace{-1.0cm}\includegraphics[width=8cm]{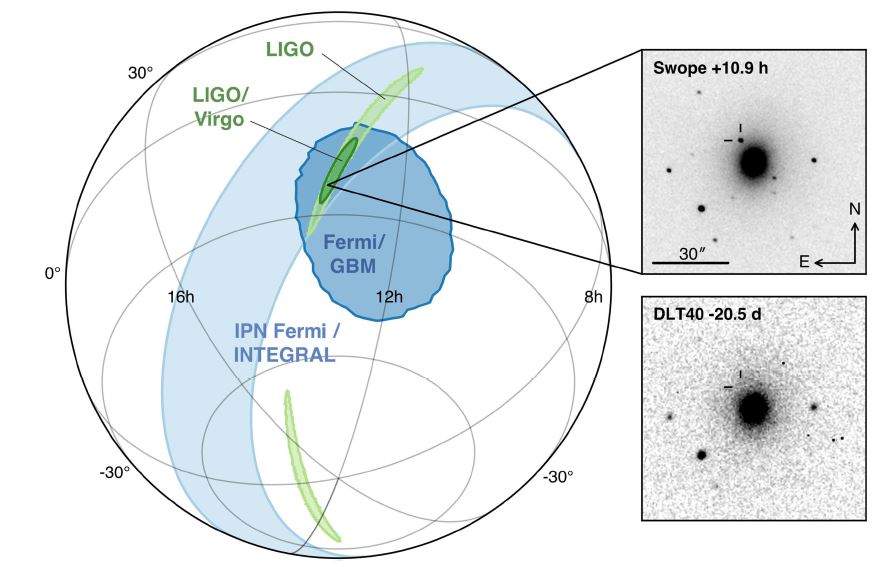}
			\hspace{-0.2cm}\includegraphics[width=5.5cm]{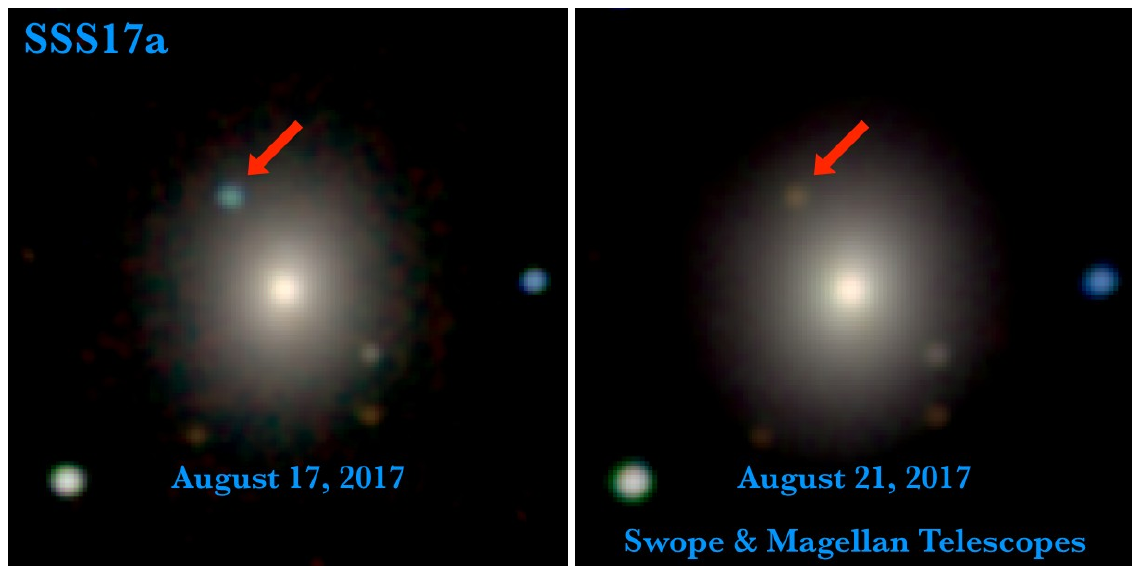}
		\end{tabular}
		\caption{Boites d'erreur sur la localisation fournies par les différents instruments, gravitationnels LIGO seul et LIGO/Virgo, et en rayons $\gamma$, Fermi et Integral (à gauche). Les images optiques au centre montrent qu'il n'y avait pas d'objet visible à cet endroit 20 jours avant la fusion (image du bas). \`A droite, emission visible et proche infrarouge. L'objet est bleu le 17 Août, puis rougit les jours suivants.}
		\label{}
	\end{center}
\end{figure}

Quasiment au même moment (1,7 seconde plus tard) un sursaut électromagnétique $\gamma$ est détecté par les satellites Fermi et Integral. L'information est automatiquement transmise à la communauté astronomique qui mobilise un grand nombre de télescopes pour essayer d'en trouver la source et de rechercher des contreparties dans les différentes longueurs d'onde (optique, IR, X, radio). L'analyse conjointe des données des interféromètres gravitationnels LIGO et Virgo conduit à la première localisation de la source et à une estimation de sa distance, environ 40\,Mpc. Cela permet aux observatoires de focaliser leur recherche dans une boite d'erreur d'environ 30 degrés carrés (qui contient une cinquantaine de galaxies) et de découvrir un objet transitoire en association avec la galaxie NGC 4993, qui est rapidement identifié à une ``kilonova''. 

Le concept de kilonova provient d'un modèle théorique décrivant l'explosion résultant de la coalescence de deux étoiles à neutrons [33]. Bien que l'étoile à neutrons soit constituée essentiellement de neutrons, sa croûte contient des empilements de noyaux lourds comme le fer ou le nickel, se répartissant selon leur densité en une structure cristalline. Certains de ces noyaux lourds ont été synthétisés lors de l'explosion en supernova qui a donné lieu à la formation de l'étoile à neutrons. Lors de la fusion des deux étoiles la majeure partie de la masse s'effondre pour former le trou noir. Mais une petite fraction des neutrons sont éjectés et se combinent alors avec les noyaux lourds existants pour former des noyaux beaucoup plus lourds, très riches en neutrons et instables. Ceux-ci se désintègrent alors rapidement, et peuvent former des noyaux moins lourds mais stables, en général bien au-delà du fer. C'est un processus de nucléosynthèse connu, dit ``$r$'' pour rapide, qui se produit dans un environnement riche en neutrons. Dans la kilonova, l'énergie émise provient de la désintégration radioactive des noyaux, et est environ 1000 fois supérieure à celle d'une nova, d'où son nom.
\begin{figure}[t]
	\begin{center}
		\includegraphics[width=9cm]{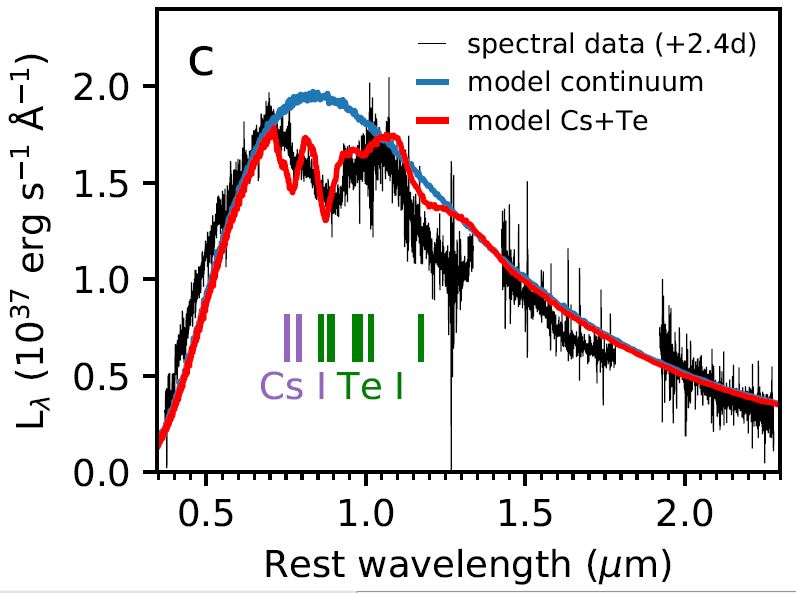}
	\end{center}
	\caption{Le spectre de la kilonova. C'est essentiellement un spectre de corps noir, avec une température de l'ordre de 6000 degrés, mais avec des raies d'absorption dues au Tellure et au Césium, deux éléments lourds (numéros atomique 52 et 55) dont la formation est difficile à expliquer dans les explosions de supernovas. Les raies sont élargies par effet Doppler dû à la vitesse d'éjection d'environ 60 000 km/s.}\label{}
\end{figure}
\begin{figure}[t]
	\begin{center}
		\includegraphics[width=10cm]{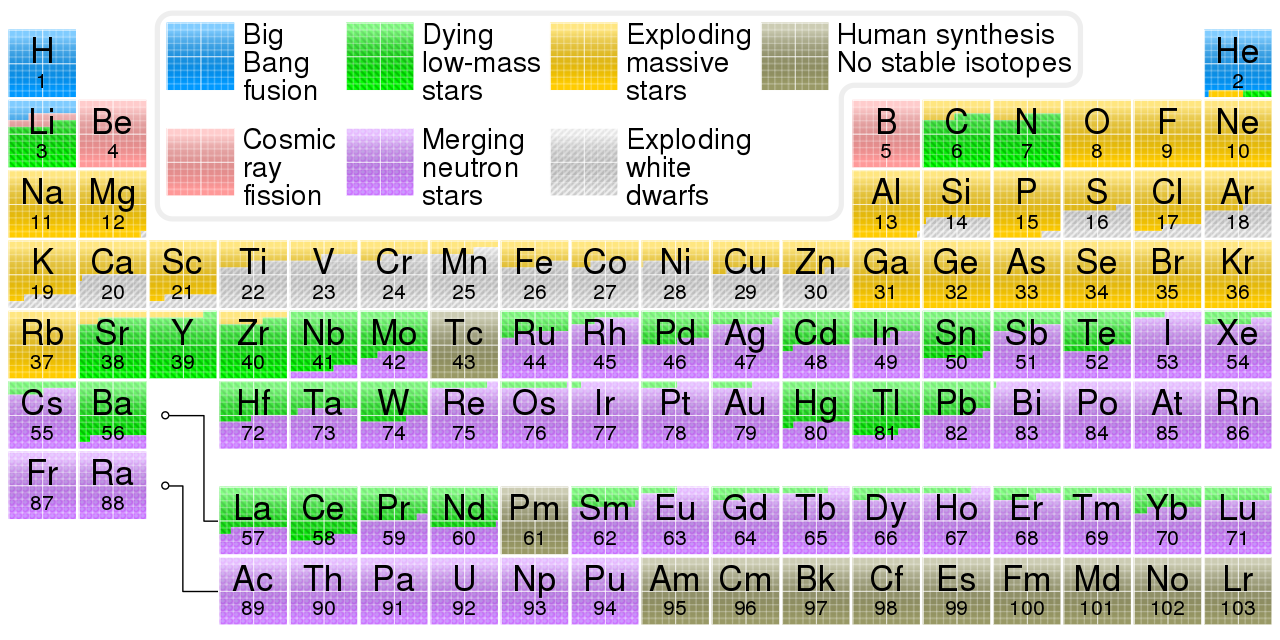}
	\end{center}
	\caption{La table des éléments avec leur origine probable. Les éléments les plus légers tels que l'hélium sont synthétisés dans le Big-Bang. Pour beaucoup d'éléments plus lourds comme le fer, on invoque divers mécanismes d'explosions de supernovas. La formation des éléments les plus lourds comme le césium, l'or, les lanthanides, l'uranium, etc. est maintenant expliquée par les coalescences d'étoiles à neutrons (code-couleur violet).}\label{}
\end{figure}

C'est une découverte majeure et la réponse à une question vieille de plus de 50 ans\,: les coalescences d'étoiles à neutrons et l'explosion cataclysmique associée constituent un site important de production d'éléments tels que l'or, le platine, l'uranium. La modélisation détaillée des réactions nucléaires dans la kilonova et la confrontation aux observations permet de valider ce modèle, et d'étudier les proportions d'éléments lourds qui sont synthétisés lors des coalescences d'étoiles à neutrons.

A noter que cette synthèse des éléments lourds dans les coalescences d'étoiles à neutrons est rendue possible par la relativité générale\,! Sans elle, pas d'ondes gravitationnelles donc pas de rapprochement possible des deux étoiles à neutrons et pas de coalescence, donc pas de synthèse de ces éléments lourds. Quelle magnifique application pratique de la relativité générale\,!\footnote{L'application pratique de la relativité générale qui est souvent citée est pour la précision du GPS qui nécessite de prendre en compte le décalage gravitationnel des fréquences.}

La différence de 1,7s entre les temps d'arrivée de l'onde gravitationnelle et du sursaut $\gamma$ est hautement significative. Elle s'explique très probablement par la différence des instants d'émission des deux types d'ondes au niveau de la source. Elle s'interprète dans le cadre d'un modèle de sursaut $\gamma$, comme le temps mis par l'énergie électromagnétique pour s'échapper, à partir d'un jet de matière relativiste émis perpendiculairement au plan orbital de la collision.

Ce délai de $1,7\,\text{s}$ confirme que la vitesse de propagation des ondes gravitationnelles est égale à celle de la lumière, à moins que $10^{-15}$ près. Ce simple fait (bien sûr en parfait accord avec la relativité générale) permet d'éliminer des théories alternatives de la gravitation, qui ont été notamment proposées pour expliquer l'énergie noire en cosmologie. Ainsi, la classe des théories ``tenseur-scalaire'' de Horndeski [34], dans lesquelles un degré de liberté scalaire est couplé de façon générale à la métrique de l'espace-temps, est fortement contrainte par cette observation.

Une propriété remarquable des ondes gravitationnelles est que l'on mesure directement avec le signal  la distance de la source [35]. Cette distance, combinée avec la mesure du décalage spectral (redshift) dans les éventuelles contreparties X ou optique, permet d'avoir une mesure directe du paramètre de Hubble-Lemaître $H_0$, c'est-à-dire le taux local d'expansion de l'Univers. Dans le cas de la fusion d'étoiles à neutrons, GW170817, avec une distance de  40 Mpc et une vitesse de récession de la galaxie NGC 4993 de 3017 km/s (après correction du mouvement propre local), on obtient une valeur de $H_0$ de 70 km/sec/Mpc. L'incertitude actuellement est cependant encore élevée, de l'ordre de 15\%. 
\begin{figure}[h]
	\begin{center}
		\includegraphics[width=9cm]{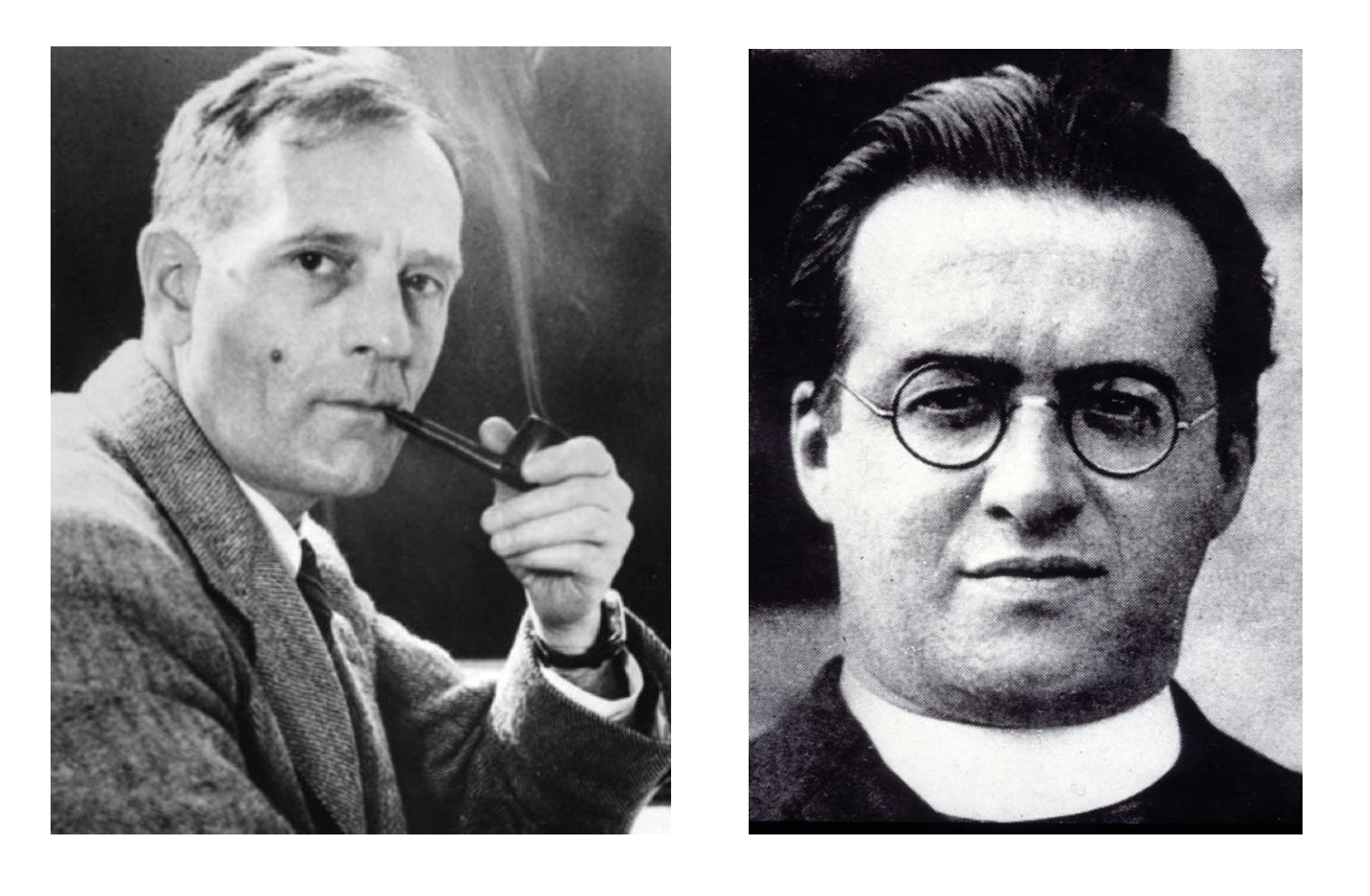}
	\end{center}
	\caption{En 2018 une résolution de l'Union Astronomique Internationale recommande de nommer ``loi de Hubble-Lemaître'' la loi d'expansion de l'Univers et le paramètre d'expansion associé $H_0$.}\label{}
\end{figure}

Un problème actuel important pour la cosmologie est la ``tension'' sur la valeur de la constante de Hubble-Lemaître, entre la valeur mesurée localement grâce à la luminosité des supernovas Ia, qui constituent des ``chandelles standard'' et dont la distance est calibrée avec des mesures de paralaxe et des étoiles céphéides, et la valeur mesurée à grand redshift en cosmologie par l'ajustement global du modèle standard aux fluctuations du fond diffus cosmologique. Cette tension est statistiquement significative. Elle signale peut-être une nouvelle physique comme une modification du modèle standard de la cosmologie aux petites échelles, ou des biais astrophysiques non compris actuellement. La valeur de $H_0$ sera de toutes façons déterminée à terme avec précision par les observatoires gravitationnels. Ici, plus besoin de calibration, c'est la relativité générale qui calibre automatiquement la distance de la source\,! 
\begin{figure}[t]
	\begin{center}
		\includegraphics[width=10cm]{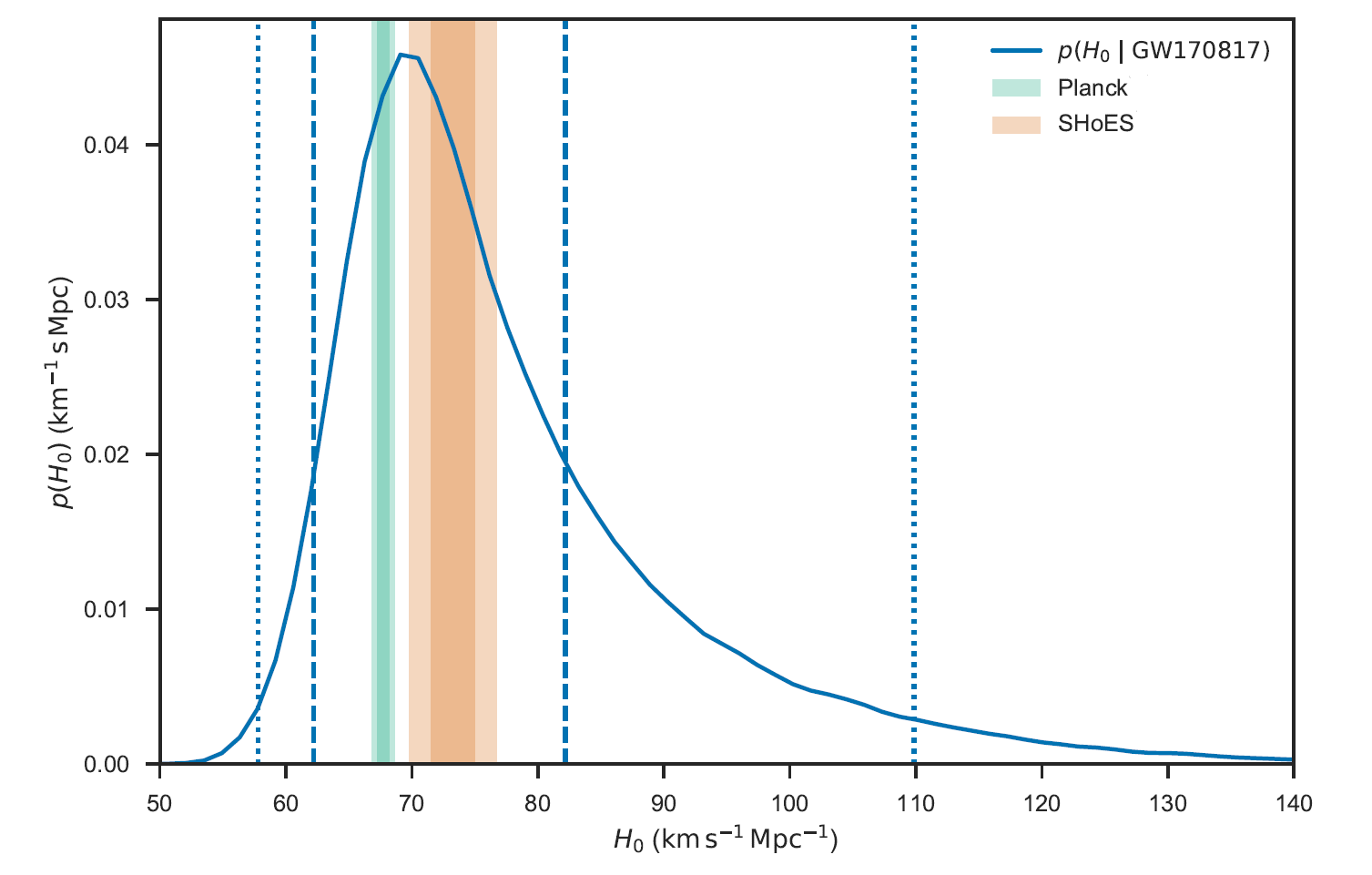}
	\end{center}
	\caption{La mesure du paramètre de Hubble-Lemaître grâce à l'évènement d'onde gravitationnelle GW170817 et au redshift mesuré de la galaxie NGC 4993 associée à la source. L'estimation la plus probable se situe entre la valeur mesurée localement avec les supernovae (expérience SHoES) et la valeur déduite des observations cosmologiques (satellite Planck).}\label{}
\end{figure}

\section{La relativité générale à l'épreuve des grandes échelles}

Nous disposons aujourd'hui d'un excellent modèle de la cosmologie, dit standard ou de concordance, basé sur la relativité générale et une hypothèse d'isotropie et d'homogénéité de l'Univers. Ce modèle est en conformité avec tout un ensemble de données observationnelles et décrit avec grande précision l'Univers aux grandes échelles, comme cela a été montré avec les mesures de fluctuations du rayonnement du fond diffus cosmologique par le satellite Planck. Dans ce modèle, la matière ordinaire dont sont constitués les étoiles, le gaz, les galaxies, etc. (essentiellement sous forme baryonique) ne forme que 4\% de la masse-énergie totale, ce qui est déduit de la nucléosynthèse primordiale des éléments légers et des mesures du fond diffus cosmologique. Nous savons aussi qu'il y a 23\% de matière noire froide (cold dark matter, CDM) sous forme non baryonique et dont nous ne connaissons pas la nature. Et les 73\% qui restent sont sous la forme d'une mystérieuse énergie noire, mise en évidence par le diagramme de Hubble des supernovae de type Ia, et dont on ignore l'origine à part qu'elle pourrait être sous la forme d'une constante cosmologique. Le contenu de l'univers à grandes échelles est donc donné par un ``camembert'' dont 96\% nous est inconnu\,!
\begin{figure}[h]
	\begin{center}
		\includegraphics[width=7cm]{figures/camembert.jpg}
	\end{center}
	\caption{Le camembert de la distribution en masse-énergie de l'Univers.}\label{}
\end{figure}

La matière noire doit être postulée dans le membre de droite des équations d'Einstein en plus de la matière baryonique, alors qu'on n'a toujours pas détecté de particule de matière noire, ni directement par des expériences en laboratoire, ni indirectement en astronomie par des observations qui seraient expliquées par l'annihilation possible de la matière noire en d'autres particules. Le meilleur candidat pour la particule de matière noire est le WIMP (Weakly Interacting Massive Particle), une particule massive qui n'interagit avec la matière ordinaire que \textit{via} l'interaction faible, mais n'a malheureusement pas été détectée dans les accélérateurs de particules tels que le Large Hadron Collider (LHC). Il y a donc actuellement un conflit majeur sur le problème de la matière noire entre le modèle de concordance cosmologique et le modèle standard de la physique des particules.

Quant à l'énergie noire, elle apparaît comme un milieu de densité d'énergie constante au cours de l'expansion, ce qui implique une violation des ``conditions d'énergie'' habituelles avec une pression négative. Sous sa forme la plus simple l'énergie noire n'est autre que la fameuse constante cosmologique d'Einstein $\Lambda$. En principe celle-ci est permise donc doit bien être incluse dans les équations de la relativité générale. Le problème survient lorsque l'on essaye de comprendre la valeur mesurée de cette constante\footnote{Elle vaut environ $10^{-123}$ en unités de Planck.} en utilisant la théorie quantique des champs. Comme l'a montré Sakharov en 1968, l'énergie des fluctuations quantiques du vide en théorie des champs prend nécessairement la forme d'une constante cosmologique. C'est pourquoi cette constante est souvent assimilée à l'énergie du vide. En théorie des champs on ignore généralement l'énergie du vide car on ne s'intéresse qu'à des différences d'énergie et que celle-ci disparaît dans la différence. Mais en relativité générale, d'après le principe d'équivalence, toutes les formes d'énergies sont la source du champ gravitationnel, et l'énergie du vide n'est qu'une forme d'énergie particulière qu'il faut donc mettre dans le membre de droite des équations d'Einstein. Lorsque l'on fait le calcul en théorie des champs de l'énergie du vide, on trouve une valeur très grande, qui n'a rien à voir avec la valeur très petite de la constante cosmologique $\Lambda$ observée. Ce problème ruine l'interprétation de l'énergie noire en tant qu'énergie du vide.

En plus des problèmes liés à l'existence et l'interprétation du secteur noir, le modèle standard est aussi confronté à des problèmes observationnels, principalement à l'échelle des galaxies [36]. Par exemple, on observe une forte corrélation entre la distribution de matière baryonique et le champ gravitationnel dans les galaxies, alors que l'on s'attendrait plutôt à une corrélation entre la distribution de matière noire et le champ, puisque la matière noire est censée dominer la masse des galaxies. Mentionnons aussi la fameuse loi empirique de Tully-Fisher [37] qui relie la masse baryonique des galaxies à leur vitesse de rotation, et qui n'est reproduite par le modèle standard qu'au prix d'ajustements fins dans la physique complexe de la matière ordinaire dans les simulations de formation des galaxies [38].

Est-il possible que l'observation de la présence de matière noire soit en fait une illusion et que l'on soit témoin d'une modification de la relativité générale\,? On pourrait penser que la modification devrait avoir lieu à grande échelle. Mais on observe des grandes galaxies elliptiques qui sont pratiquement dépourvues de matière noire, alors qu'elle domine dans des galaxies naines. Or une formule étonnante appelée MOND (MOdified Newtonian Dynamics) indique plutôt une modification dans un régime de champs de gravitation plus faibles qu'une certaine échelle d'accélération caractéristique dénotée $a_0$. Cette formule due à Milgrom [39] reproduit la mystérieuse corrélation observée entre la présence de matière noire dans les galaxies et l'échelle d'accélération (loi de Milgrom), ainsi que les courbes de rotation plates pour les galaxies et la loi de Tully-Fisher. Contrairement à ce que son nom laisse penser, MOND n'est pas une théorie de gravitation modifiée, mais c'est juste une formule empirique qui rend compte de façon très simple de tout un ensemble de faits observationnels au niveau des galaxies. L'échelle caractéristique de MOND vaut $a_0 \sim 1.2 \,10^{-10} \text{m}/\text{s}^2$. 

La formule MOND constitue peut-être une clé pour comprendre le problème de la matière noire. Est-ce que MOND aurait le potentiel de défier la relativité générale\,? En principe oui, mais malgré beaucoup d'efforts aucune théorie relativiste satisfaisante modifiant la relativité générale n'a été trouvée, qui rende compte à la fois de la formule MOND à l'échelle des galaxies et incorpore les succès du modèle standard à l'échelle cosmologique. 

La relativité générale reste notre chère théorie de la gravitation\,!

\vspace{0.5cm}
\centerline{{\bf Références}}
\vspace{0.5cm}

\begin{enumerate}
\item Bracco, C., and Provost, J.P., \textit{``Poincaré and relativity: the logic of the 1905 Palermo Memoir''}, in the thirteenth Marcel Grossmann Meeting on General Relativity, p. 2045 (2015).
\vspace{-0.2cm}
\item Laue, M.V., \textit{``The Entrainment of Light by Moving Bodies According to the Principle of Relativity''}, Annalen der Physik {\bf 23}, 989 (1907).
\vspace{-0.2cm}
\item Poincaré, H., Comptes Rendus Acad Sci. Paris {\bf 140}, 1504 (1905).
\vspace{-0.2cm}
\item Einstein, A., Grossmann, M., \textit{``Entwurf einer verallgemeinerten Relativitätstheorie und einer Theorie der Gravitation''}, BG Teubner (1913).
\vspace{-0.2cm}
\item Abuter, R., et al. (GRAVITY collaboration), Astron. and Astrophys. {\bf 636}, L5 (2020).
\vspace{-0.2cm}
\item Blanchet, L., Hébrard, G., and Larrouturou, F., \textit{``Detecting the general relativistic orbital precession of the exoplanet HD80606b''}, Astron. and Astrophys. {\bf 628}, A80 (2019).
\vspace{-0.2cm}
\item Shapiro, I.I., \textit{``Fourth test of general relativity''}, Phys. Rev. Lett. {\bf 789}, 13 (1964).
\vspace{-0.2cm}
\item Will, C.M., \textit{``Theory and experiments in gravitational physics''}, Cambridge U. Press (1993).
\vspace{-0.2cm}
\item Lorentz, H.A., and Droste, J., \textit{``The motion of a system of bodies under the influence of their mutual attraction, according to Einstein's theory''}, reprinted p. 330, Nijhoff (1937).
\vspace{-0.2cm}
\item Einstein, A., Infeld, L., and Hoffmann, B., \textit{``The Gravitational Equations and the Problem of Motion''}, Ann. Math. {\bf 39}, 65 (1938).
\vspace{-0.2cm}
\item Eisenstaedt, J., \textit{``Einstein et la relativité générale''}, CNRS editions (2002).
\vspace{-0.2cm}
\item Eötvös, R.V., Math. nat. Ber. Ungarn. {\bf 8}, 65 (1890).
\vspace{-0.2cm}
\item Roll, P.G., Krotkov, R., and Dicke, R.H., \textit{``The equivalence of inertial and passive gravitational mass''}, Ann. Phys. (N.Y.) {\bf 442}, 26 (1964).
\vspace{-0.2cm}
\item Touboul, P., Métris, G., Rodrigues, M., et al., \textit{``MICROSCOPE mission: first results of a space test of the equivalence principle''}, Phys. Rev. Lett. {\bf 23}, 231101 (2017).
\vspace{-0.2cm}
\item Brillet, A., and Hall, J., \textit{``Improved laser test of the isotropy of space''}, Phys. Rev. Lett. {\bf 42}, 549 (1979).
\vspace{-0.2cm}
\item Pound, R.V., and Rebka, G.A., \textit{``Apparent weight of photons''}, Phys. Rev. Lett. {\bf 4}, 337 (1960).
\vspace{-0.2cm}
\item Vessot, R.F.C., and Levine, M.W., \textit{``A test of the equivalence principle using a space-borne clock''}, Gen. Rel. and Grav. {\bf 10}, 181 (1979).
\vspace{-0.2cm}
\item Cacciapuoti, L., and Salomon, C., \textit{``Space clocks and fundamental tests: the ACES
experiment''}, Eur. Phys. J. Spec. Top. {\bf 57}, 127 (2009).
\vspace{-0.2cm}
\item Williams, J., Turyshev, S., and Boggs, D., Phys. Rev. Lett. {\bf 93}, 261101 (2004).
\vspace{-0.2cm}
\item Carter, B., Phys. Rev. Lett. {\bf 26}, 331 (1971).
\vspace{-0.2cm}
\item Kerr, R., \textit{``Gravitational field of a spinning mass as an example of algebraically special metrics''}, Phys. Rev. Lett. {\bf 11}, 237 (1963).
\vspace{-0.2cm}
\item Genzel, R., Eisenhauer, F., and Gillessen, S., Rev. Mod. Phys. {\bf 82}, 3121 (2010).
\vspace{-0.2cm}
\item Einstein, A., \textit{``Näherungsweise Integration der Feldgleichungen der Gravitation''}, Sitzber. Preuss. Akad. Wiss. Berlin {\bf 1}, 688 (1916).
\vspace{-0.2cm}
\item Einstein, A., \textit{``Uber Gravitationswellen, Sitzber}. Preuss. Akad. Wiss. Berlin, 154 (1918).
\vspace{-0.2cm}
\item Hulse, R., and Taylor, J., \textit{``Discovery of a pulsar in a binary system''}, Astrophys. J. {\bf 195}, L51 (1975).
\vspace{-0.2cm}
\item Taylor, J., and Weisberg, J., \textit{``A new test of general relativity: gravitational radiation and the binary pulsar PSR 1913+16''}, Astrophys. J. {\bf 253}, 908 (1982).
\vspace{-0.2cm}
\item Peters, P.C., and Mathews, J., \textit{``Gravitational Radiation from Point Masses in a Keplerian Orbit''}, Phys. Rev. {\bf 131}, 435 (1963).
\vspace{-0.2cm}
\item Damour, T., and Deruelle, N., \textit{``Radiation reaction and angular momentum loss in small angle gravitational scattering''}, Phys. Lett. A, 81–84 (1981).
\vspace{-0.2cm}
\item Abbott, B., et al., LIGO-Virgo scientific collaboration, \textit{``Observation of gravitational waves from a binary black hole merger''}, Phys. Rev. Lett. {\bf 116}, 061102 (2016).
\vspace{-0.2cm}
\item Blanchet, L., \textit{``Gravitational radiation from post-Newtonian sources and inspiralling compact binaries''}, Living Rev. Rel. {\bf 17}, 2 (2014).
\vspace{-0.2cm}
\item Buonanno, A., and Damour, T., \textit{``Effective one-body approach to general relativistic two-body dynamics''}, Phys. Rev. D {\bf 59}, 084006 (1999).
\vspace{-0.2cm}
\item Abbott, B., et al., LIGO-Virgo scientific collaboration, \textit{``GW170817: Observation of Gravitational Waves from a Binary Neutron Star Inspiral''}, Phys. Rev. Lett. {\bf 119}, 161101 (2017).
\vspace{-0.2cm}
\item Metzger, B., et al., Monthly Not. R. Astron. Soc. {\bf 406}, 2650 (2010).
\vspace{-0.2cm}
\item Horndeski, G.W., \textit{``Second-order scalar-tensor field equations in a four-dimensional space''}, Int. J. Theor. Phys. {\bf 10}, 363 (1974).
\vspace{-0.2cm}
\item Schutz, B., Nature {\bf 323}, 310 (1986). 
\vspace{-0.2cm}
\item Blanchet, L., and Famaey, B., \textit{``La relativité générale à l'épreuve des grandes échelles''}, in La Recherche, Hors-Serie {\bf 16}, 70 (2015). 
\vspace{-0.2cm}
\item Tully, R.B., and Fisher, J.R., \textit{``A new method of determining distances to galaxies''}, Astron. Astrophys. {\bf 54}, 661 (1977).
\vspace{-0.2cm}
\item Famaey, B., and McGaugh, S., \textit{``Modified Newtonian dynamics (MOND): Observational phenomenology and relativistic extensions''}, Living Rev. Relativ. {\bf 15}, 10 (2012).
\vspace{-0.2cm}
\item Milgrom, M., \textit{``A modification of the Newtonian dynamics as a possible alternative to the hidden mass hypothesis}. Astrophys. J. {\bf 270}, 365 (1983).
\end{enumerate}


\end{document}